\documentclass[aps,pra,twocolumn,superscriptaddress,groupedaddress]{revtex4}
\usepackage{graphicx}
\usepackage{dcolumn}
\usepackage{bm}       
\usepackage{amssymb} 
\usepackage{amsmath}
\usepackage{amsfonts}
\usepackage{amsthm}
\usepackage{xfrac}
\usepackage[cp1252]{inputenc}
\usepackage{epsf}
\usepackage{lpic}
\usepackage{color}
\newcommand{\be}{\begin{equation}}
\newcommand{\ee}{\end{equation}}
\newcommand{\ba}{\begin{align}}
\newcommand{\eal}{\end{align}}
\newcommand{\dg}{ ^{\dagger}}
\newcommand{\hml}{\mathcal{H}}
\newcommand{\dpp}{\partial}
\newcommand{\la}{\langle}
\newcommand{\ra}{\rangle}

\begin{document}

\title{Quantum signatures in quadratic optomechanics}
\author{J. D. P. Machado}
\author{R.J. Slooter}
\author{Ya. M. Blanter}
\affiliation{Kavli Institute of Nanoscience, Delft University of Technology, Lorentzweg 1, 2628 CJ Delft, The Netherlands}

\date{\today}

\begin{abstract}
We analyze quantum effects occurring in optomechanical systems where the coupling between an optical mode and a mechanical mode is quadratic in displacement (membrane-in-the-middle geometry). We show that it is possible to observe quantum effects in these systems without achieving the single-photon strong coupling regime. We find that zero-point energy causes a mechanical frequency shift, and we propose an experimental way to measure it. Further, we show that it is possible to determine the phonon statistics from the cavity transmission, and propose a way to infer the resonator's temperature based on this feature. For completeness, we revisit the case of an isolated system and show that different types of mechanical quantum states can be created, depending on the initial cavity state. In this situation, mechanical motion undergoes collapse and revivals, and we compute the collapse and revival times, as well as the degree of squeezing. 
\end{abstract}

\maketitle

\section{Introduction}

Creation and detection of quantum states of mechanical motion have become an increasing topic of interest due to their key role in tests of fundamental physics \cite{romero}. An ongoing effort in the field of optomechanics is to reach regimes where the coupling between a mechanical resonator and an optical cavity mode enables quantum-state engineering. Despite the variety of observed physical phenomena, direct observations or creation of nontrivial mechanical quantum states have not yet been achieved. One difficulty hindering these realizations in optomechanical systems is the nature of the radiation pressure coupling. For most experimental realizations, the coupling is linear in displacement, which is not suitable to directly measure the energy of the mechanical resonator \cite{medindo}, nor to allow the creation of nontrivial quantum states for the resonator via an unconditional evolution from Gaussian states \cite{exact}. It has been proposed that mechanical quanta could be directly observed if the coupling were quadratic in displacement \cite{phononjumps,quasiMIM}. As the interaction between light and mechanical motion depends on the system's geometry, the coupling can be made quadratic by placing a membrane at a symmetric point of a cavity mode (node/anti-node).

 Quadratic couplings have already been implemented in membrane-in-the-middle setups \cite{MIMtrivial,Karuza,MIM1}, ultracold atoms \cite{fio} and levitating dielectric particles \cite{levitatednanoparticles}. Due to the direct dependence on the phonon number, this type of coupling is particularly suited to observe quantum jumps of the phonon number \cite{phononjumps,quasiMIM}, as well as to characterize the phonon statistics by a direct measurement of the cavity spectrum \cite{quasiMIM}. Other quantum features associated with this type of coupling are phonon shot noise \cite{medindo}, antibunching \cite{unico}, and squeezing \cite{nunnen,amitrai,batacharia}. Squeezing and antibunching reveal the potential for the creation and manipulation of mechanical quantum states.

 Although this type of interaction enables many interesting effects, its nonlinear nature has hampered extensive theoretical analyses of the quantum behaviour arising in this regime. Theoretical approaches have focused on the resonant form of the interaction \cite{phononjumps} together with adiabatic elimination of the optical mode \cite{unico,nunnen}, or linearized dynamics \cite{maismembs}. However, for an undriven cavity, an exact diagonalization for this quadratic coupling is possible \cite{amitrai,batacharia}, as well as an extension including the coupling to the environment \cite{nori}. The exact solution reveals that the time-evolution of the isolated system naturally modifies the mechanical state, enabling the creation of squeezed states and states with a star shaped Wigner function \cite{batacharia}. Another consequence of the quadratic coupling is the photon state dependence of the mechanical frequency, which leads to collapse and revivals of the mechanical motion for coherent cavity states \cite{amitrai}, analogous to cavity QED \cite{cummings}. Collapse and revivals have been extensively studied \cite{cummings,redencao,polacos,picanha,pimairux,gandacolapso,russo+polacos} and it was shown that a revival's envelope is characteristic of the quantum state \cite{polacos,gandacolapso,russo+polacos}, leading to quantum state reconstruction methods based on the shape of the revivals \cite{endoscopia}.

In this article, we explore the full potential of the quadratic coupling and show that it enables quantum effects never discussed previously, such as the shift of the mechanical frequency due to zero-point motion and the characterization of the phonon statistics via the cavity transmission. For completeness, we first review the isolated system case in Sec. \ref{sec:isolee}, but provide previously unknown results as well. In subsec. \ref{sec:colapso}, we compute the mechanical displacement and variance for a cavity coherent state, show that the mechanical motion undergoes collapse and revivals for coherent and squeezed cavity states, and obtain expressions for the revival and collapse times for the first time. In subsec. \ref{sec:estado}, we compute the time-evolution of the mechanical resonator's state, and show that a variety of mechanical quantum states can be created depending on the initial cavity state. In particular, we find that mechanical superposition-like states can be created.

 Another interesting peculiarity of the quadratic coupling is the effect of zero-point energy (ZPE). It has been suggested that ZPE leads to an observable Casimir force \footnote{It has been remarked that ZPE does not couple directly to matter, and that ZPE-like effects are an asymptotic limit of microscopic models \cite{hereges}. The scope of this manuscript is not to discuss the nature of this frequency shift, but rather to propose an experimental scheme to measure it.} in optomechanical systems in the quadratic coupling regime \cite{levitatednanoparticles}, and an experimental proposal to measure this effect based on the phase shift of a probe beam has been advanced \cite{medo}. In Sec. \ref{sec:zpe}, we propose an alternate scheme based on our findings that ZPE leads to a mechanical frequency shift, which is enhanced by the presence of multiple quadratically coupled optical modes. Our proposal consists of measuring the mechanical frequency at different cavity points and compare with the calculated frequency shifts. This type of dynamical force measurement has the advantage of using a static cavity and it is expected to be more precise than phase shift or amplitude shift measurement techniques \cite{casimirmede}. Further, we believe that this proposal can be tested with the current technology \cite{MIM1, fio}, and analyze its feasibility.

Apart from enabling the creation of nontrivial quantum states, the quadratic coupling also enables the identification of the mechanical state. Besides the observation of mechanical quanta through quantum jumps \cite{phononjumps}, it is also possible to directly determine the phonon statistics. This feature is possible due to the form of the resonant interaction, which enables a quantum non-demolition measurement of the phonon number. In Sec. \ref{sec:driven}, we compute the cavity transmission for a weak probe laser, and show that the mechanical state affects the transmission. In the single-photon strong coupling regime, the transmission profile enables the direct determination of the phonon statistics, because the cavity transmission exhibits well-resolved peaks for each phonon Fock state, whose relative height corresponds to the probability of finding the resonator in that particular Fock state. Even outside this regime, it is possible to distinguish between different mechanical states by analyzing the transmission profile.  For a coherent state, the transmission presents a Gaussian-like shaped profile, whereas for a thermal state, the cavity transmission is asymmetric with a tail governed by the Boltzmann distribution, and we propose a way to determine the resonator's temperature based on this feature.
Finally, we present the conclusions and discuss the range of applicability of our results in Sec. \ref{sec:conclusions}.

\section{Isolated system}
\label{sec:isolee}

An optomechanical system can be modeled as two coupled harmonic oscillators (the mechanical resonator and the optical cavity). When the mechanical element is placed at a node (or anti-node) of a cavity mode, the reflection symmetry ensures that the coupling is quadratic in displacement, for displacements much smaller than the cavity length. If the mechanical element is a linear dielectric, the coupling is proportional to the cavity field intensity, and the Hamiltonian is given by \cite{phononjumps,MIMtrivial,unico}
\be
\hml=\omega_ca\dg a +\Omega b\dg b +g \Big(a\dg a +\frac{1}{2}\Big)(b\dg +b)^2\,,
\label{eq:hamil}
\ee
where $\omega_c$ and $\Omega$ are respectively the cavity and mechanical frequencies, $g$ the coupling constant, and $a$($b$) the photon (phonon) annihilation operator. As the interaction preserves the photon number, Eq. (\ref{eq:hamil}) represents a quadratic form for the phonon operators, whose eigenfrequencies and eigenstates depend on the photon number. Thus, Eq.(\ref{eq:hamil}) can be diagonalized via a photon number dependent squeezing operator
\be
S(\hat{r}(a\dg a))= \sum_{n=0}^{\infty} e^{\frac{1}{2}r(n)( (b\dg)^2-b^2)}|n\rangle \langle n| \,,
\label{eq:squeezor}
\ee
where $|n\rangle$ refers to the photon Fock state. Using the short-hand notation $\hat{\chi}=g(a\dg a+\frac{1}{2})$ and choosing $r(n)$ to be real, the action of the squeezing operator defined in Eq.(\ref{eq:squeezor}) on Eq.(\ref{eq:hamil}) leads to the diagonalized Hamiltonian
\be
\hml_{D}=\omega_ca\dg a+ \Big(\Omega\cosh (2\hat{r})+2\hat{\chi}e^{2\hat{r}}\Big)\big(b\dg b+\frac{1}{2}\big)\,.
\label{eq:hdiag}
\ee
Eq.(\ref{eq:hdiag}) is obtained by imposing that the nondiagonal terms vanish (the ones with $bb$ and $b\dg b\dg$), which implies
\be
 (\Omega+2\hat{\chi})\sinh (2\hat{r})+2\hat{\chi}\cosh(2\hat{r})=0\,,
\ee
and determines the squeezing parameters to be (cf. \cite{batacharia})
\be
r(n)=-\frac{1}{4}\log\left(1+\frac{4g(n+\sfrac{1}{2})}{\Omega}\right)\,.
\label{eq:raa}
\ee
Combining Eqs.(\ref{eq:hdiag}) and (\ref{eq:raa}) leads to (cf. \cite{amitrai})
\be
\hml_{D}=\omega_ca\dg a+\sqrt{\Omega^2+4g\Omega(a\dg a+\sfrac{1}{2})}\Big(b\dg b+\frac{1}{2}\Big)\,.
\label{eq:Hdiagonal}
\ee
It is seen from Eq.(\ref{eq:Hdiagonal}) that the resonant term of Eq.(\ref{eq:hamil}) is the first order term in $g$ of the energy spectrum, and that for $2g n\geq \Omega$, higher order terms overcome the resonant term. This marks the point where the rotating-wave approximation (RWA) fails, and the full nonlinearity must be considered. With this transformation, it is possible to evaluate the time-evolution of the quantum state for arbitrary coupling strengths, as well as any physical observable, via the time-evolution operator $W(t)=S\dg (\hat{r}(a\dg a))e^{-i\hml_Dt}S(\hat{r}(a\dg a))$. 

\subsection{Collapses and revivals of mechanical motion}
\label{sec:colapso}

\begin{figure}[h]
\includegraphics[scale=0.55]{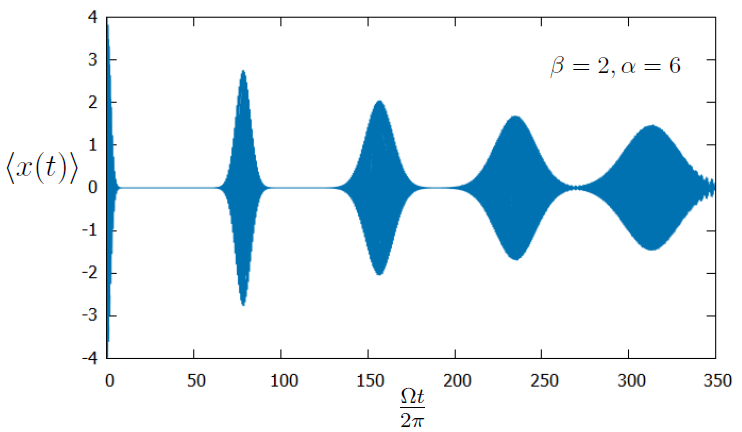}
\caption{Expected mechanical displacement for the initial coherent states $|\beta=2, \alpha=6\rangle$, and coupling $g=0.01\Omega$. The mechanical displacement rapidly decays to 0 (the collapse event) and after $\approx78$ periods, the oscillation reappears (the revival). The revivals become smaller and broader with time, until interference between successive revivals occurs.\label{fig:collrev1}}
\end{figure}

It can be seen from Eq.(\ref{eq:Hdiagonal}) that the oscillating frequency of each oscillator depends on the quantum state of the other. This quantum state dependence influences the time-evolution of the mechanical displacement $x(t)$, given by
\be
x(t)=W\dg(t)x(0)W(t)=\cos \big(\varpi t \big)x(0) +\frac{\Omega}{\varpi} \sin \big(\varpi t \big)p(0)\,,
\label{eq:xt}
\ee
where $\varpi = \sqrt{\Omega^2+4g\Omega(a\dg a+\frac{1}{2})}$, $x=b+b\dg$, and $p=i(b\dg -b)$. As $\varpi$ depends on the photon number, $x(t)$ has a different frequency for each $|n\rangle$, with a relative weight dependent of $|\langle n|\psi\rangle|^2$, where $|\psi\rangle$ is the cavity state. Thus, measuring the resonator's spectral density provides a direct way to directly determine the photon statistics in the optical domain. A consequence of this photon number sensitivity for the mechanical displacement is that these frequencies interfere, leading to collapse and revivals of mechanical motion. This interference makes the resonator's mean displacement quickly drop to 0 (the collapse), only to reappear again at a latter time (the revival) \cite{amitrai}. This behaviour is displayed in Fig. \ref{fig:collrev1} for the initial coherent states $|\beta=2, \alpha=6\rangle$, where $\beta$ ($\alpha$) is the phonon (photon) state.

\begin{figure}[h]
\includegraphics[scale=0.5]{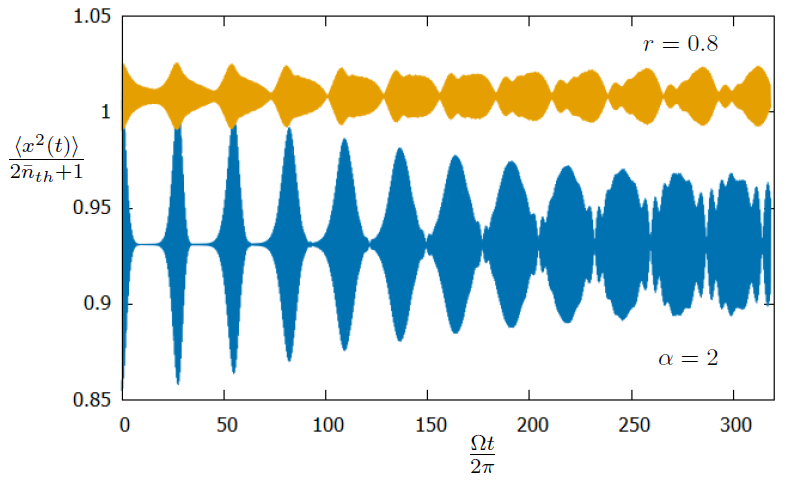}
\caption{Collapse and revival behaviour of the displacement variance for an initial thermal state of the resonator, and for the cavity states $|\alpha=2\rangle$ and $|r=0.8\rangle$, where $\alpha$ and $r$ denote coherent and vacuum squeezed states, respectively. The shape of the revivals is characteristic of the cavity state, and for a vacuum squeezed state, the revival's envelope changes with time.\label{fig:collrev2}}
\end{figure}

In general, the mechanical motion is not periodic because of the incommensurability of the frequencies (see Eq.(\ref{eq:Hdiagonal})). Consequently, each revival is smaller and broader than the previous one, and after several revivals, these start to overlap and interfere with each other. From that moment on, the motion exhibits a seemingly chaotic behaviour. The collapse and revival times can be estimated using well-known techniques from cavity QED \cite{redencao,pimairux,gandacolapso}. For large coherent photon states ($|\alpha|^2\gg1$), the Poissonian distribution can be approximated by a Gaussian distribution, and for $g \ll \Omega$, $\varpi$ can be expanded in powers of $g$. Replacing the sum in the Fock basis by an integral via the Poisson summation formula, the mechanical displacement can be expressed as an oscillation whose amplitude is modulated by Gaussian envelops of the form
\ba
\langle x(t) \rangle\approx  2\beta \sum_m &\cos(\varpi_\alpha t+2\pi |\alpha|^2 m)\nonumber\\
&\times \text{exp}\left(-\frac{1}{2}\bigg(\frac{t-m T_{rev}}{T_{coll}}\bigg)^2\right)\,,
\end{align}
where $\alpha$ is the initial cavity coherent state, the initial phonon coherent state $\beta$ was taken to be real, and $\varpi_\alpha= \sqrt{\Omega^2+4g|\alpha|^2\Omega}$. The collapse and revival times ($T_{coll}$ and $T_{rev}$, respectively) are given by
\be
T_{rev}=2\pi|\alpha|T_{coll}=\frac{\pi \sqrt{\Omega^2+4g|\alpha|^2\Omega}}{g\Omega}\,.
\label{eq:momentos}
\ee
In contrast to cavity QED \cite{redencao,pimairux,gandacolapso}, the revival time in these optomechanical systems depends on the average photon number, and for high photon numbers ($4g\Omega |\alpha|^2\gg 1$), the collapse time becomes independent of the mean photon number ($T_{coll}=(g\Omega)^{-\sfrac{1}{2}}$).  This collapse and revival behaviour is not restricted to the displacement, and it is visible for any mechanical observable as long as $2\pi |\alpha|>1$. For example, the displacement variance for a thermal phonon state and a coherent state $\alpha$ for the cavity is
\begin{align}
\langle x^2(t)\rangle\approx & (2\bar{n}_{th}+1)\Bigg(\frac{\Omega+2g|\alpha|^2}{\Omega+4g|\alpha|^2}\nonumber\\
&-e^{-|\alpha|^2}\sum_{n=0}^{+\infty}\frac{|\alpha|^{2n}}{n!}\frac{2g n}{\Omega+4g n}\cos (2\varpi_n t)\Bigg)\,,
\label{eq:squ}
\end{align}
where $\bar{n}_{th}$ is the average phonon number of the thermal state. As seen from Fig. \ref{fig:collrev2} and Eq.(\ref{eq:squ}), the interaction creates squeezing, but for an initial coherent cavity state, the non-oscillating quadrature uncertainty can never be reduced to less than a half of the thermal uncertainty. Nevertheless, at the peak of each revival, the minimum uncertainty has no lower bound. Further, Fig. \ref{fig:collrev2} shows that the appearance of collapse and revivals is not restricted to coherent states, and that the temporal envelope is characteristic of the cavity state. Particularly, for a vacuum squeezed photon state $|r\rangle$, the shape of the revival is elongated due to the superposition of the revival's echoes \cite{gandacolapso}, with an envelope which changes in time.

\subsection{Evolution of the quantum state}
\label{sec:estado}

\begin{figure}[h]
\centering
\includegraphics[scale=0.35]{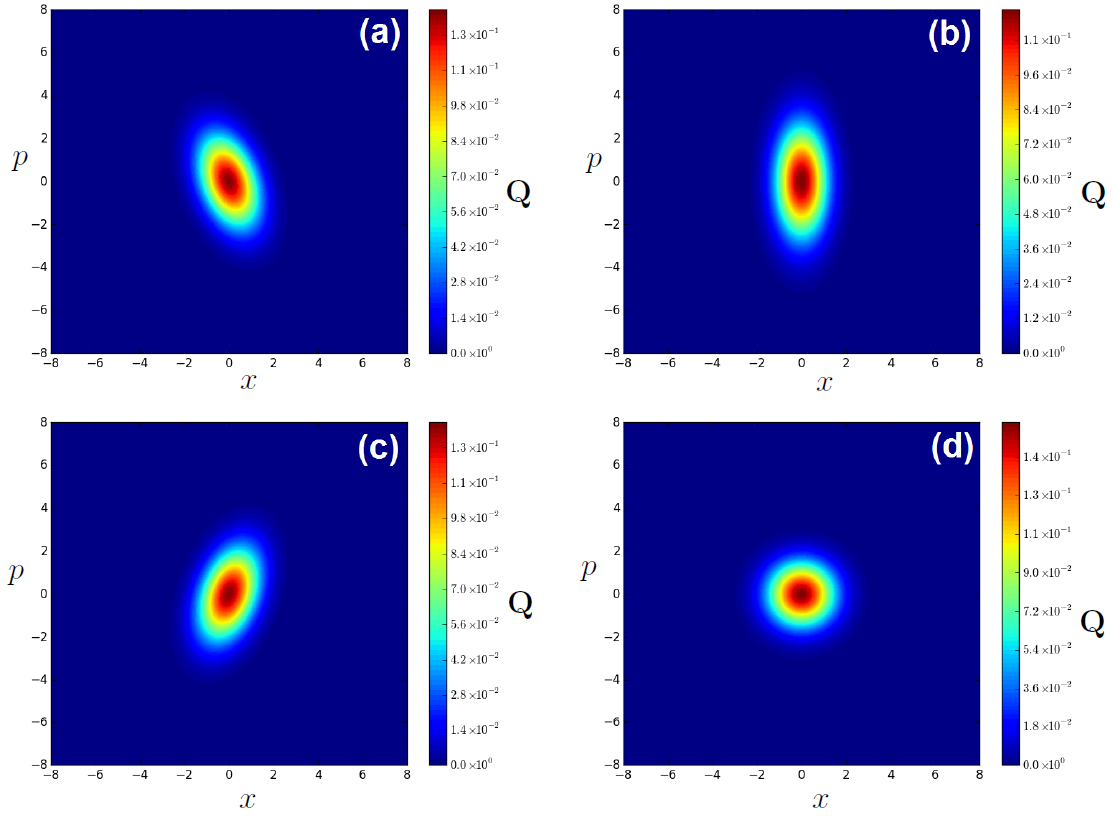}
\caption{Time-evolution of Q-function of the mechanical state for $g=0.01\Omega$, an initial cavity Fock state $N=100$, and mechanical ground-state, after $(\sfrac{1}{4},\sfrac{1}{2} ,\sfrac{3}{4},1)$ effective mechanical periods (panels (a),(b),(c), and (d), respectively). The interaction produces a periodic squeezing of the resonator.\label{fig:states}}
\end{figure}

As shown in Eq.(\ref{eq:squ}), the resonator experiences squeezing, and so the quantum state does not remain static. The time-evolution of the mechanical state for the initial phonon state $|0\rangle$ and cavity Fock state $|100\rangle$ is displayed in Fig. \ref{fig:states}, where the Husimi Q-function for the resonator was computed using QuTiP \cite{qutip}. For an initial cavity Fock state, the mechanical state is periodically squeezed, with the period determined by the effective phonon frequency in Eq.(\ref{eq:xt}). Note that this feature solely depends on the initial cavity (Fock) state. The reason for this effect comes from the interaction form (see Eq.(\ref{eq:hamil})), which comprises a photon-number dependent mechanical frequency, and squeezing interaction.\par

\begin{figure}[h]
\centering
\includegraphics[scale=0.53]{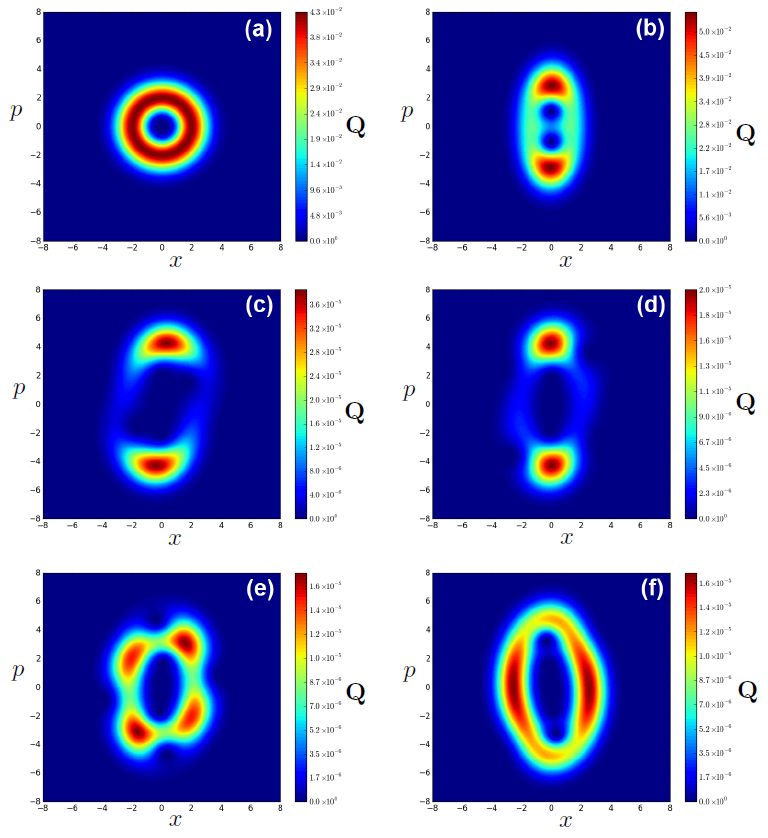}
\caption{Time-evolution of Q-function of the mechanical state for $g=0.01\Omega$, an initial phonon Fock state $n=2$, and cavity coherent state $\alpha=\sqrt{40}$, after $(0, 1\sfrac{1}{2}, 130, 260, 260\sfrac{1}{4}, 261)$ effective mechanical periods (panels (a),(b),(c),(d),(e) and (f), respectively). After $1\,\sfrac{1}{2}$ periods, the Fock state (a) suffers a quadrature squeezing (b). Several periods later, the state evolves into a superposition-like state (c). The mechanical state undergoes rapid transformations within a period. This is seen from the quantum state in (d) evolving to a state resembling a superposition of four coherent states (e), and afterwards to a seemingly distorted Fock state (f). \label{fig:morestates}}
\end{figure}

Besides squeezed states, which have already been discussed in the literature \cite{batacharia}, other interesting quantum states can be created, depending on the initial cavity state. For a coherent cavity state $|\alpha=\sqrt{40}\,\rangle$ (and initial mechanical Fock state $|n=2\rangle$), the mechanical state evolves into a quantum state resembling a superposition state after several periods (Fig. \ref{fig:morestates}, panels (c,d)). The state undergoes rapid changes even within a period. As seen in Fig. \ref{fig:morestates}, the quantum state in panel (d) evolves into a state resembling a superposition of four coherent states (panel(e)), and afterwards to a seemingly distorted Fock state (panel (f)).\par
\par
Note that even though the existence of two peaks in the $Q$ function is insufficient to say that the resonator is in a superposition state, it is clear from panel (d) of Fig. \ref{fig:morestates} that the quantum state is not a statistical mixture of coherent states.
For the statistical mixture of coherent states $\rho=\frac{1}{2}(|i\beta\ra \la i\beta |+|-i\beta\ra \la -i\beta |)$, the Q function is
\be
Q_{mixt}(\alpha)=\frac{1}{2\pi}(e^{-|\alpha-i\beta|^2}+e^{-|\alpha+i\beta|^2})\,,
\ee
which has two separate peaks like the aforementioned figure, but it is also Gaussian distributed along the line Im$\{\alpha\}=p=0$. On the other hand, panel (d) of Fig. \ref{fig:morestates} displays a small deep around the origin and it has two maxima along the $p=0$ line, and so the state cannot be a simple statistical mixture. The overall quantum state is quite complex because the mechanical state is entangled with the light state, but since no dissipation nor noise are present, this state is expected to display quantum correlations.

\section{Zero-point energy effects}
\label{sec:zpe}

The analysis above considered only the coupling between one optical mode and the resonator. However, even if there are no photons for a given cavity mode, Eq.(\ref{eq:Hdiagonal}) reveals that this optical mode still plays a role due to ZPE. As the interaction preserves the photon number, the multi-mode case can be easily approached for an arbitrary number of modes. With the substitution $g(a\dg a+\sfrac{1}{2}) \to \sum_j g_j (a_j\dg a_j+\sfrac{1}{2})$, the transformation in Eq.(\ref{eq:squeezor}) can be generalized to diagonalize the multi-mode Hamiltonian. Even though the coupling $g$ is small for most physical implementations of quadratic coupling, the contribution of several cavity modes enhances the mechanical frequency shift produced by ZPE.

 A simple way to measure this frequency shift is to place the membrane at a high symmetry point of the cavity (such as the centre of the cavity, where all optical modes couple quadratically to the membrane) and measure its frequency $\Omega_{centre}$, and then shift the membrane to a point of low symmetry (such as close to one of the end mirrors) and measure the frequency $\Omega_{end}$ at this position. Note that the cavity should not be driven to prevent undesired contributions. Although it is not possible to monitor the membrane's position if the optics couples quadratically to the mechanics, the mechanical frequency can still be determined by a laser probe out of axes and independent of the cavity system or a mechanical probe. In the multi-mode single-photon weak-coupling regime, ZPE is responsible for the frequency difference 
\be
\Omega_{end}-\Omega_{centre}\approx 2\sum_{j, even} g_j=G\,.
\label{oxala}
\ee
It is possible to implement this proposal with the existing technology \cite{MIMtrivial,Karuza,MIM1}, and it represents an alternative to force or displacement measurements of ZPE. In order for this scheme to be feasible, the frequency difference $G$ must surpass the mechanical linewidth $\Gamma$, which does not require achieving the single-photon strong coupling regime. So far, $G$ has never been determined, and the enhancement produced by all the even cavity modes is yet unknown. Apart from relatively high values for cold atoms implementations \cite{fio}, the quadratic coupling for a single mode is in general quite small ($\sim 5\mu$Hz \cite{Karuza}). Although mechanical linewidths on the order of a few $\mu$Hz exist \cite{minigama}, such small frequency differences may be difficult to detect. However, G is expected to increase by a few orders of magnitude with the use of low frequency resonators ($g\propto x_{ZPM}^2\propto \Omega^{-1}$), and at least by another 2 orders of magnitude with the use of highly reflective membranes \cite{vitali}. The combination of these improvements may bring the quadratic single-photon coupling to the Hz regime, where it can surpass existent mechanical linewidths, and ensure the feasibility of this proposal.\par

\section{Driven cavity}
\label{sec:driven}

\begin{figure}[h]
\includegraphics[scale=0.5]{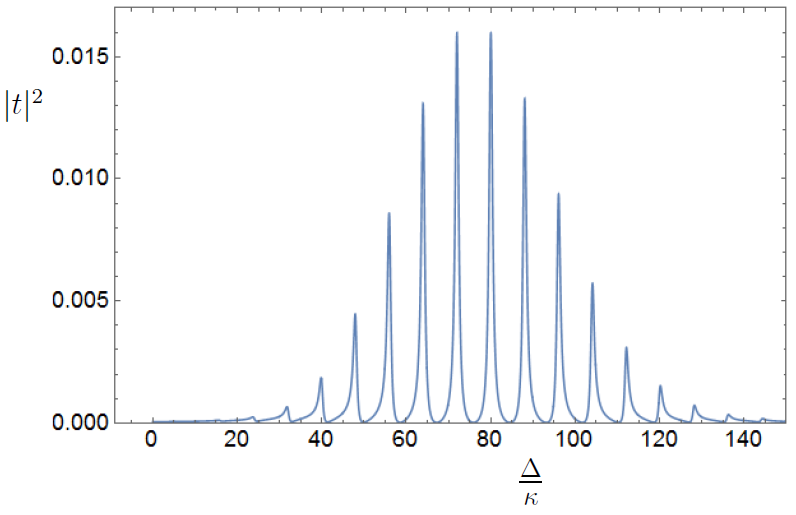}
\caption{Cavity transmission for $g=4\kappa$ and a coherent mechanical state with $|\beta|^2=10$. In the single-photon strong coupling regime, the relative height of the $n^{th}$ peak corresponds to the probability to find the mechanical resonator in a Fock state $|n\ra$.
\label{fig:transparenciaideal}}
\end{figure}

 An ubiquitous obstacle hindering the observation of quantum effects in optomechanical systems is dissipation. As the cavity decay rate $\kappa$ can easily surpass the coupling strength, after the short time-scale of $\kappa^{-1}$, the light stored inside the cavity will have leaked out and it will no longer interact with the mechanical resonator. To ensure a steady photon number, the cavity must be driven. For weak driving (i.e. $|\alpha|^2\ll g^{-1}\Omega$, with $|\alpha|^2$ the intracavity photon number created by the probe laser), the off-resonant interaction term $a\dg a (bb+b\dg b\dg)$ is small and can be disregarded (see Sec.\ref{sec:isolee}). Thus, using RWA and including a driving term in the Hamiltonian, the effective Hamiltonian in the drive reference frame is
\be
\hml=-\Delta a\dg a+\Omega b\dg b+i\mathcal{E}(a-a\dg)+2ga\dg a b\dg b\,,
\ee
where $\Delta=\omega_L-\omega_c-g$ is the detuning from the laser frequency $\omega_L$ and $\mathcal{E}$ the driving strength. Within RWA, the interaction preserves the phonon number, which enables a quantum non-demolition measurement of the phonon number. To take the effects of dissipation into account, we make use of a Fokker-Planck equation for the Husimi functions $Q_n(\alpha)=\frac{1}{\pi}\la n,\alpha|\rho|n,\alpha\ra$, where $n$ refers to a phonon Fock state, $\alpha$ to a photon coherent state, and $\rho$ to the density matrix.
Using standard master equation techniques \cite{charmichael}, the Fokker-Planck equation governing the behaviour of the system is
\begin{eqnarray}
&\dpp_tQ_n =i(-\Delta+2gn)(\alpha\dpp_\alpha-\alpha^*\dpp_{\alpha^*})Q_n+\kappa\dpp_\alpha\dpp_{\alpha^*}Q_n\nonumber\\
&+\kappa Q_n+\frac{\kappa}{2}(\alpha\dpp_\alpha+\alpha^*\dpp_{\alpha^*})Q_n-\mathcal{E}(\dpp_\alpha+\dpp_{\alpha^*})Q_n \ .
\label{eq:Qevolucion}
\end{eqnarray}
After the short timescale $\kappa^{-1}$, the cavity reaches its equilibrium state, and so we are interested only on the stationary properties and not in the dynamics. Additionally, we assume that the resonator is in a stationary state. This can be achieved by waiting until the steady-state is reached and optically probing the resonator afterwards. Even if the resonator is not in a  steady-state, the stationary situation can still be considered if the mechanical thermalization rate $\Gamma \bar{n}_{th}$ is much smaller than the inverse of the measurement time. The steady-state solution of (\ref{eq:Qevolucion}) is
\be
Q_{n,ss}=\frac{1}{\pi}exp\left(-\Big|\alpha-\frac{\mathcal{E}}{\frac{\kappa}{2}-i(\Delta-2gn)}\Big|^2\right)\,,
\ee
which leads to the intracavity field amplitude
\be
\langle a\rangle_{ss}=\int \alpha Tr[Q_{n,ss}]d^2\alpha =\sum_n \frac{\mathcal{E}p_n}{\frac{\kappa}{2}-i(\Delta-2gn)}\,,
\label{eq:ass}
\ee
where $p_n$ is the probability to find the mechanical resonator in the Fock state $|n\ra$. Eq.(\ref{eq:ass}) features a set of peaks, each corresponding to a specific phonon number, and with a relative height of $p_n$ (the probability to find the mechanical resonator in a given Fock state $n$). An interesting consequence of this photon-phonon interaction is thus the fingerprint left by the phonon statistics on the cavity field. It is then possible to determine the phonon statistics by measuring the cavity transmission $|t|^2$. The transmission is defined as the ratio between the coherent output power and the coherent input power
\be
|t|^2=\left|\frac{\langle a_{out}\rangle}{\langle a_{in}\rangle}\right|^2=\left|\frac{\kappa_e\langle a\rangle}{2\mathcal{E}}\right|^2\,,
\label{eq:transmision}
\ee
where $\kappa_e$ is the decay rate through the output mirror (taken to be $\approx \kappa$ onwards). If the single-photon strong coupling is reached ($4g\gg \kappa$), each of the Fock peaks in the transmission is well-resolved and the phonon statistics can be immediately identified (see Fig. \ref{fig:transparenciaideal}). The determination of the phonon number is not possible with the standard radiation pressure interaction $a\dg a (b+b\dg)$, and it is a unique feature of the quadratic coupling.

\begin{figure}[h]
\includegraphics[scale=0.5]{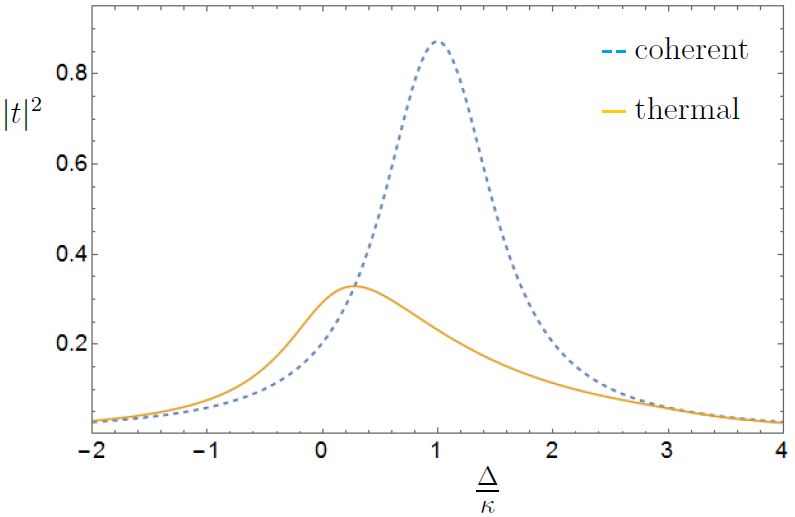}
\caption{Comparison between the cavity transmission for a mechanical thermal state (thick yellow line) and for a coherent state (blue dashed line), with $g=0.01\kappa$ and with an average of 50 phonons for both states. In this regime, the quantum state is not fully resolved, but the transmission lineshape still exhibits distinctive features characteristic of the mechanical state.\label{fig:comptransmit}}
\end{figure}

Although the single-photon strong coupling regime is far from being achieved experimentally, there are still interesting features outside this regime. Particularly, one can still characterize the state without the peaks being fully resolved. In the multi-phonon strong coupling regime (defined here as 4$g\bar{n}_b\gg \kappa$, with $\bar{n}_b$ the average phonon number), the transmission exhibits a lineshape characteristic of the mechanical state, as shown in Fig. \ref{fig:comptransmit}.
Even if the average phonon number is the same, the transmission for a coherent state differs substantially from the transmission for a thermal state. The phonon number distribution for a large coherent state resembles a Gaussian, which is directly reflected in the transmission profile (see Fig. \ref{fig:comptransmit} and note the deviation from $\Delta=0$), whereas for a thermal state, there is a strong asymmetry in the transmission. This asymmetry is a consequence of the Boltzmann exponential trend, and it can be used to determine the temperature of the resonator. For a thermal state, $p_n=\frac{1}{\bar{n}_{th}}\Big(\frac{\bar{n}_{th}}{\bar{n}_{th}+1}\Big)^{n+1}$, and using Eqs. (\ref{eq:ass},\ref{eq:transmision}), the transmission is found to be
\ba
|t|^2&=\frac{\kappa_e^2}{\kappa^2+4\Delta^2}\nonumber\\
&\times \left| _2F_1 \bigg(1, -\frac{\kappa i+2\Delta}{4g}, 1 - \frac{\kappa i+2\Delta}{4g},\frac{\bar{n}_{th}}{\bar{n}_{th}+1}\bigg)\right|^2
\label{eq:termometro}
\end{align}
where $_2F_1(a,b,c,z)$ is the Hypergeometric function. This transmission thermometry provides a simple method to determine the resonator temperature with the use of Eq.(\ref{eq:termometro}). This asymmetry is in principle visible with a slight improvement on the state-of-the-art setups since for $g\sim 100\mu$Hz and for kHz resonators at room temperature, the multi-phonon strong coupling regime can be achieved for cavity linewidths up to the MHz range.

Although we are mostly concerned about the properties of the mechanical resonator, the latter can also be used to change the cavity properties. As the area below the transmission lineshape is independent of the quantum state, and higher $\bar{n}_b$ lead to  broader and smaller transmission profiles, the mechanical element can be used as switch to control the light exiting the cavity.

\section{Conclusions}
\label{sec:conclusions}
Summarizing, we analyzed the quantum features present in optomechanical systems with a coupling quadratic in displacement. For the isolated system case, we have shown the possibility to create nontrivial mechanical quantum states, such as superposition-like states. Although the interaction is able to modify the mechanical state in a nontrivial way, the same does not occur for the cavity state. As the interaction preserves the photon number, the photon statistics remain unchanged. We have also shown that collapse and revivals of mechanical motion occur in these systems due to the photon state dependence of the mechanical frequency, and we calculated the characteristic collapse and revival times, as well as the degree of squeezing. Further, we computed the mechanical frequency shift induced by ZPE, and proposed a way to measure this shift by placing the membrane at different points of the cavity. For the case when the cavity is weakly probed, we have shown that the cavity transmission can be used to identify the phonon statistics, and proposed a method to determine the resonator's temperature based on the cavity transmission profile. Both this transmission thermometry and the ZPE mechanical frequency shift can be measured without reaching the single-photon strong coupling regime, and we expect that they can be experimentally tested with the current technology.
\\

Acknowledgments: We thank Jo\~{a}o Moura and Clemens Sch\"{a}fermeier for helpful discussions and suggestions, and the Dutch Science Foundation (NWO/FOM) for its financial support.\\

\end{document}